\documentclass[aps,pre,showpacs,twocolumn,groupedaddress,floatfix]{revtex4}

\usepackage{graphicx}
\usepackage{amssymb}
\usepackage{amsmath}
\usepackage{color}
\usepackage{psfrag}
\usepackage{times}

\begin{document}

\title{Network of recurrent events for the Olami-Feder-Christensen model}


\author{Tiago P. Peixoto}
\email[]{tpeixoto@if.usp.br}
\affiliation{Instituto de F\'{i}sica, Universidade de S\~{a}o Paulo,
Caixa Postal 66318, 05315-970 - S\~{a}o Paulo - Brazil}
\author{J\"orn Davidsen}
\email[]{davidsen@phas.ucalgary.ca}
\affiliation{Complexity Science Group, Department of Physics \& Astronomy,
University of Calgary, Calgary, Alberta T2N 1N4, Canada }

\date{\today}

\begin{abstract}
We numerically study the dynamics of a discrete spring-block model
introduced by Olami, Feder and Christensen (OFC) to mimic
earthquakes and investigate to which extent this simple model is
able to reproduce the observed spatiotemporal clustering of
seismicty. Following a recently proposed method to characterize
such clustering by networks of recurrent events [Geophys. Res.
Lett. {\bf 33}, L1304, 2006], we find that for synthetic catalogs
generated by the OFC model these networks have many non-trivial
statistical properties. This includes characteristic degree
distributions --- very similar to what has been observed for real
seismicity. There are, however, also significant differences
between the OFC model and earthquake catalogs indicating that this
simple model is insufficient to account for certain aspects of the
spatiotemporal clustering of seismicity.
\end{abstract}


\maketitle


\section{Introduction\label{sec:intro}}

Describing and modelling the spatiotemporal organization of
seismicity and understanding the underlying physical mechanisms of
earthquake triggering have been proven challenging. Inspired by
empirical regularities which include self-similar scaling laws
like the Omori law and the Gutenberg-Richter law
\cite{turcotte,stein}, a wealth of mechanisms and models have been
proposed including the concepts of a critical earthquake, of
self-organized criticality, and more generally of the seismogenic
crust as a self-organized complex system requiring a so-called
system approach (see, for example, Refs.~\cite{mulargia,rundle03}
for a review). Yet, the origin of the observed non-trivial
emergent features of earthquake occurrence is still one of the
main unresolved problems in the field.

Resolving this issue may require measuring the microscopic state
variables  --- the stress and strain at every point within the
earth along active earthquake faults --- and their exact dynamics.
This is currently impossible. However, the associated
earthquake patterns are readily observable making macroscopic
approaches based on the concept of \emph{spatiotemporal point
processes} feasible, where the description of each earthquake is
reduced to its size or magnitude, its epicenter and its time of
occurrence. Describing the patterns of seismicity may shed light
on the fundamental physics since these patterns are emergent
processes of the underlying many-body nonlinear system.

Recently, such an approach has led to the identification of new
properties of the clustering of seismicity in space and time
\cite{bak02,helmstetter03b,corral04,davidsen04,baiesi04,
  shcherbakov04,davidsen05m,helmstetter05,livina05,molchan05,
  shcherbakov06,felzer06}.  In particular, the observed spatiotemporal
clustering of seismicity suggests that the usual mainshock/aftershock
scenario --- where each event has at most one correlated predecessor of
larger magnitude --- is too simplistic and that the causal structure of
seismicity could extend beyond immediately subsequent events. To
quantify such correlations, a general procedure for inferring a
plausible causal structure from clusters of localized events has been
introduced in Refs.~\cite{davidsen05pm,davidsen06pm}. The approach
allows one to study the dynamical organization of spatiotemporal
activity in terms of the topology of complex networks
~\cite{albert02,newman03} and has led to the detection of unexpected
statistical features for earthquake catalogs from California. Most
importantly, the approach provides a new and independent estimate of the
rupture length and its scaling with magnitude.

In this article, we investigate to which extent the simple dynamics of
the Olami-Feder Christensen (OFC) model \cite{olami92} is sufficient to
account for the observed spatiotemporal clustering of seismicity as
characterized by the above network approach. The OFC model is maybe the
simplest model in the class of self-organized critical (SOC) models
which exhibit a phenomenology resembling seismicity. This includes the
aforementioned Gutenberg-Richter law for the frequency-magnitude
distribution ~\cite{olami92,christensen92b} and the Omori law for
aftershocks \cite{hergarten02,helmstetter04} as well as some statistical
properties of epicenter locations and dynamics~\cite{peixoto06}.  In
addition, the OFC model is of special relevance in the context of
SOC. Non-equilibrium systems are called self-organized critical if they
reach a stationary state characterized by power laws and the absence of
characteristic scales --- without the need for fine-tuning an external
parameter such as the temperature~\cite{bak,jensen}. This is typically
the case for slowly driven systems with fast avalanche-like dissipation
events.  Unlike the paradigmatic SOC sandpile model~\cite{bak87}, where
any amount of dissipation is enough to introduce a scale which breaks
criticality, the existence of criticality in the OFC model with
dissipation is still
debated~\cite{carvalho00,lise01,lise01b,miller02,boulter03,wissel06}.

We show that when the causal structure of the OFC model is studied
with the network method mentioned above, it does indeed reveal some
similarities to seismicity, such as the degree distributions of the
network, and some aspects of the recurrence time and distance
statistics. However, there are several important differences 
--- including the absence of a rupture length --- that
severely hinder its adoption as a plausible (sole) description of the
fundamental mechanisms responsible for the spatiotemporal correlations
of seismicity. Many of those discrepancies seem to be closely related
to the fact that the model is defined on a regular, discrete
lattice, and to the existence of quasi-periodic attractors
\cite{peixoto06}. These features, which are totally absent in
seismicity, manifest themselves quite clearly in several aspects of
the recurrence network. Also,
the recurrence time statistics show a characteristic time due to the 
fact that the system is uniformly driven. 

The outline of our paper is as follows. In Section~\ref{model}, we
review the OFC model, and Section~\ref{method} summarizes the
network method to search for signs of causal structure in
spatiotemporal data. Section~\ref{results} presents the results
obtained for the OFC model, which are compared to real
seismicity. We conclude in Section~\ref{summary}.

\section{Olami-Feder-Christensen model\label{model}}

The OFC model~\cite{olami92} is inspired by the Burridge-Knopoff
spring-block model~\cite{burridge67}, and is defined on a square
lattice of size $L^2$. To each site $(i,j)$, a ``tension''
$z_{ij}$ is assigned, initially chosen at random from the interval
$[0,z_c[$. The entire system is driven slowly, with every $z_{ij}$
increasing uniformly. Whenever a site reaches the threshold
tension $z_{ij} = z_c$, a relaxation event --- called ``avalanche'' or
``earthquake'' --- starts. The originating site of the avalanche
$(i,j)$ is called the epicenter. The dynamics of such an event is
as follows: A fraction $\alpha$ of the tension at the epicenter is
transferred to each of its four neighbors $z_{i\pm 1,j\pm 1} =
z_{i\pm 1,j\pm 1} + \alpha z_{ij}$, and the tension at the site
itself is reset $z_{ij}=0$. If the tension at any of the
neighboring sites reaches the threshold, $z_{i\pm 1,j\pm 1} \ge
z_c$, the same toppling rules are applied again. This dynamics continues
until there are no more sites in the system with $z_{ij} \ge z_c$.
Then, the tension increases uniformly again until the next
avalanche occurs. Without loss of generality, we set $z_c=1$. The
total number of topplings during an avalanche is called the
``size'' of the avalanche $s$. In contrast, the total number of
sites that toppled (ignoring multiple topplings of the same site)
is called the ``area'' of the avalanche $a$. The time in the system can
be measured either discretely, by the number of events (``iterations''),
or continuously, by the cumulative tension injected in the system when
it is driven (the so called ``natural'' time scale). The parameter
$\alpha$ defines the level of local conservation of the system.
For $\alpha=0.25$ the system is locally conservative and for
$\alpha<0.25$ it is dissipative. We consider here only the case of
open boundary conditions, i.e., the sites at the border of the
lattice transfer tension out of the system (or to an imaginary
neighbor), so the system is always globally nonconservative.

After a transient regime, which is increasingly longer for smaller
values of $\alpha$, the system reaches a stationary state which has a
distribution of event sizes resembling a power-law in the tail for the
largest, numerically accessible system sizes (see Ref.~\cite{wissel06}
for an extensive review and the most recent results). For the values
of $\alpha \in\{0.18; 0.2; 0.22\}$ and $L=1000$ we consider here, the
power law tail has an unique exponent around
$-1.8$~\cite{lise01}. Yet, the occurrence of events is not uniformly
distributed within the lattice, and tend to happen closer to the
boundaries~\cite{lise01b,peixoto06}. This ``border effect'' prevents
the event size statistics from obeying finite-size scaling (FSS) with
the size of the lattice. Only when events close to the border are
ignored and an internal subset of the system is considered, FSS is
realized~\cite{lise01b}.  Interestingly, this border effect does not
seem to scale with system size~\cite{peixoto06}. To accommodate for
these observations, we generally restrict our analysis to synthetic
catalogs generated by the OFC model which contain only those events
involving sites which are all at a distance of at least 100 sites away
from the boundary. The effective systems size is thus $L_\text{bulk}=800$.

\begin{figure}[htbp]
\psfrag{a=0.18}[bc][bc][0.7]{$\alpha=0.18$}
\psfrag{a=0.20}[bc][bc][0.7]{$\alpha=0.20$}
\psfrag{a=0.22}[bc][bc][0.7]{$\alpha=0.22$}
\psfrag{<a_r>}[bc][bc]{$\langle a_r \rangle$}
\psfrag{a}[bc][bc]{$a$}
\psfrag{<a_r>/a}[bc][tc][0.7]{$\langle a_r\rangle/a$}
\includegraphics*[width=\columnwidth]{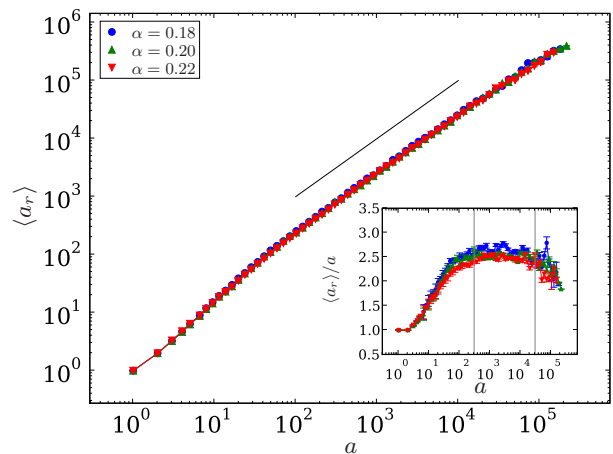}
\caption{\label{fig:rectangle-area} (Color online) The average
area $\langle a_r \rangle$ of the smallest rectangle on the
lattice containing a given avalanche of spatial extent $a$ for
different values of the dissipation parameter $\alpha$. Only
avalanches contained within the bulk, $L_\text{bulk}=800$, are
considered (see text for details). The solid line corresponds to a
linear increase. Inset: Rescaled version indicating that the area
scales linearly with $a$ independent of $\alpha$ for $300 \leq a
\leq 30000$ (area enclosed by straight lines).}
\end{figure}

An important aspect in seismicity is the effect of the detection
threshold: Earthquake catalogs are only complete above a certain
magnitude value which depends on the network of seismometers deployed
in a given area. Thus, it is crucial to understand how the statistical
properties of seismicity vary with the magnitude threshold. To
address this issue in the OFC model, we consider catalogs for
different lower threshold values of event sizes
$s_{\text{th}}$. Reasonable choices of these values are limited by the
particular dynamics of the OFC model and lattice effects. For example,
it is well known that events of size one have their own separate
dynamics and obey a different statistics \cite{wissel06}. This is easy
to justify, since events of size one will occur even if there is no
self-organization, i.e., $\alpha=0$. The effects of this particular
dynamics and of the discrete lattice can be clearly seen in the
spatial shapes of the avalanches as summarized in
Fig.~\ref{fig:rectangle-area}: The ratio between the average area
$\langle a_r \rangle$ of the smallest rectangle which contains an
avalanche of area $a$ and $a$ varies significantly with $a$ for all considered
values of $\alpha$. Here the average is taken over all events of area
$a$. Only for events above a certain size $a_{\text{th}} \sim 10^2$,
does $a_r$ scale linearly with $a$ (see inset of
Fig.~\ref{fig:rectangle-area}) --- independent of $\alpha$. Yet, this
linear scaling is limited to values of $a < 3 \times 10^4$ due to the
finite system size. Since area and size of an avalanche are basically
identical for the parameters of $\alpha$ considered here
\footnote{The area $a$ of an avalanche can in principle be
different from its size $s$, since sites can topple more than
once. However, there is a transition value of $\alpha$, below
which multiple topplings are impossible. This value can be
obtained by considering the extremal situation where a site with
tension $1$ is about to topple, whose neighbors also have tension
$1$. After the site topples, it will also cause the topplings of
the neighbors, which will then return tension to original site.
The amount of tension returned is $4\alpha(1+\alpha)$. If this
value is below $1$, then the original site will not topple a
second time, which means that no site in any other situation will
topple a second time. Thus, the transition value is
$\alpha=(\sqrt{2}-1)/2 \approx 0.207$. Below this value, $a=s$ for
any event. Simulations show that even when the value of $\alpha$
is slightly above this transition, the occurrence of multiple
topplings is negligible. For the range of $\alpha \in [0.22,0.18]$
studied here, multiple topplings play no detectable role.},
we only consider catalogs with lower size threshold 
$s_{\text{th}}\in [300;30000]$ in the following.

\begin{table}
    \caption{\label{num_events}Number of events $N$ in the catalogs generated by
      the OFC model for different values of $\alpha$ and lower size threshold
      $s_{\text{th}}$.}
    \begin{ruledtabular}
    \begin{tabular}{c|c|c|c}
    $s_{\text{th}}$  & $\alpha=0.18$ & $\alpha=0.20$ & $\alpha=0.22$  \\ \hline
    300 & 1000000  & 1000000  & 1000000 \\
    500 & 641933  &  676324  & 685237 \\
    1000  & 327124  & 394235  & 400339 \\
    3000  & 107550  & 148311  & 156189 \\
    5000  & 60041  & 87102  & 94968 \\
    10000  & 22205  & 40208  & 42991 \\
    30000  & 2547  & 8432  & 8909 \\
    \end{tabular}
    \end{ruledtabular}
\end{table}

To be more specific, each synthetic catalog generated by the OFC model contains
$N$ consecutive events or avalanches occurring after the statistically
stationary regime has been reached\footnote{The sizes of the discarded
  transients were $\sim 10^{10}$, $\sim 7\times10^9$ and $\sim 4\times10^9$
  iterations for $\alpha=0.18$, $0.2$ and $0.22$, respectively.} for $L=1000$
and $L_\text{bulk}=800$. Each event $k$ is characterized by its size $s_k$, its
epicenter location $(i,j)_k$ and its time of occurrence $t_k$ measured in terms
of the continuous natural time scale (the discrete avalanche number time scale
of the OFC model \cite{davidsen00} is basically equivalent). The various
catalogs for different values of $\alpha$ and $s_{\text{th}}$ are given in
Table~\ref{num_events}.

\section{Network of recurrences\label{method}}

We analyze the catalogs generated by the OFC model according to a
recently proposed method which has been proven helpful in characterizing
the spatiotemporal clustering of seismicity and in detecting causal
signatures between events \cite{davidsen05pm,davidsen06pm}. The
essential idea is to extend the notion of a \emph{recurrence} to
spatiotemporal point processes: An event is defined to be a recurrence
of any previous event if it occurred closer in space than all
intervening events.  Recurrences are therefore \emph{record breaking
  events} with respect to distance. By linking each event to its
recurrences, a directed network~\footnote{A network is composed of a set
  of discrete elements called vertices (or nodes) and a set of pairs of
  vertices (which can be ordered or not), called edges (or links), which
  may describe some relationship between two vertices. Two vertices are
  neighbors if there is a edge between them. The number of neighbors of
  a given vertex is called the \emph{degree} of the vertex. If the edges
  of the network are directed, then the vertex can have in-neighbours
  and out-neighbours, depending on the direction of the corresponding
  edge, and therefore also an \emph{in-degree} and an
  \emph{out-degree}. See~\cite{newman03} for an extensive and detailed
  review of these concepts.} of recurrences is obtained: Each event
$a_k$, with $k=1,...,N$, is a vertex in the network and a directed edge
from $a_k$ to $a_{m}$ exists for $k<m$ if $a_m$ is a recurrence of
$a_k$, i.e., if the distance between $(i,j)_{m}$ and $(i,j)_k$ is
smaller than the distance between $(i,j)_k$ and $(i,j)_{k'}$ for all
events $a_{k'}$ with $k<k'<m$. This definition assumes that the events
are ordered according to their occurrence in time $t_k$.  Obviously,
each recurrence or edge can be characterized by the time interval
$t=t_m-t_k$ between the two connected events $k$ and $m$ and by the
spatial distance $l$ between the two. Note that the mapping of the
dynamics to a network is entirely data-driven and does not impose any
arbitrary space and time scales other than those associated with the
given event or earthquake catalog --- in contrast, for example, to
methods typically used to define aftershocks. Comparing the statistical
properties of the network of recurrences for a given catalog to the
properties of a network obtained for a random point process without any
causal relation between events highlights relevant parts of the
underlying causal dynamical process(es) generating the pattern. For the
OFC model, such a random process or null model can be obtained by
shuffling the entire catalog, containing all events: Shuffle the sizes
and the epicenter locations separately while keeping the times of
occurrence. Later the $s_{\text{th}}$ threshold can be raised, and the
corresponding shuffled versions of each catalog can be obtained.
For a random spatiotemporal point process in continuous
space and time (CST), several statistical properties of the network of
recurrences are even known analytically~\cite{davidsen06pm}.

\section{Results\label{results}}

In the following, we do not only compare the network properties
of the OFC catalogs with those of their random or shuffled
counterparts but also with the network properties found for
earthquake catalogs from southern California
~\cite{davidsen05pm,davidsen06pm}. The particular focus is on the
network topology as summarized, for example, in 
Refs.~\cite{albert02,newman03} and on distributions associated with 
the temporal and spatial distances between an event and its recurrences.

\subsection{Topological properties\label{network}}

\subsubsection{Network growth\label{sec:growth}}

\begin{figure}[htbp]
  \psfrag{a=0.18}[bl][bl][0.7]{$\alpha=0.18$}
  \psfrag{a=0.20}[bl][bl][0.7]{$\alpha=0.20$}
  \psfrag{a=0.22}[bl][bl][0.7]{$\alpha=0.22$}
  \psfrag{a=0.18 shuffled}[bl][bl][0.7]{$\alpha=0.18$, shuffled}
  \psfrag{a=0.20 shuffled}[bl][bl][0.7]{$\alpha=0.20$, shuffled}
  \psfrag{a=0.22 shuffled}[bl][bl][0.7]{$\alpha=0.22$, shuffled}
  \psfrag{<k>}[bc][bc]{$\left<k\right>$}
  \psfrag{N}[bc][bc]{$N$}
\includegraphics*[width=\columnwidth]{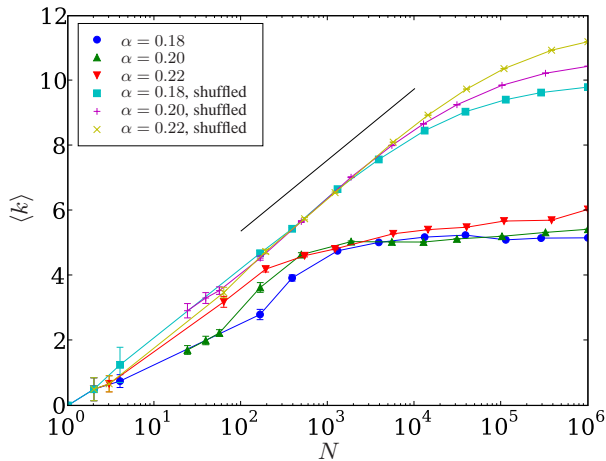}
\caption{\label{growth} (Color online) Mean degree $\langle k
\rangle$ vs. catalog size $N$ for $s_{\text{th}}=300$ and
different values of alpha, for the regular and shuffled OFC catalogs. 
$N=10^6$ corresponds to the full
catalogs --- see Table~\ref{num_events}. The solid line is of the
form $0.95 \ln(N)+ \text{const}$.}
\end{figure}

An important aspect of the analysis of networks and their dynamics
is the change in their topology with time. In the case of networks
of recurrences, this corresponds to the situation when a given
catalog is extended to cover a longer time period and, thus,
contains more events. In general, some properties of the network
of recurrences depend on its size $N$. For random CST events, 
one such property is the average degree
$\left<k\right>$ or equivalently the average number of recurrences 
per event. It grows with the size of the network as
$\left<k\right>\approx \ln(N)$ for $N\gg 1$~\cite{davidsen06pm}.
While this is roughly what we find for the shuffled OFC catalogs, 
the original OFC catalogs show quite a
different growth as can be seen in Fig.~\ref{growth}.
Independently of the value of $\alpha$, the average degree
``saturates'' quickly around a value close to $5$, and grows very
little even after the catalog is a couple of orders of magnitude
larger. When the catalog is shuffled, a portion of the growth can
be well represented by $\ln(N)$ but only up to $\sim 10^4$ events,
after which it grows more slowly.

\begin{figure}[htbp]
  \psfrag{d1}[c][c][0.7]{$\mathbf{d_1}$}
  \psfrag{d2}[c][c][0.7]{$\mathbf{d_2}$}
  \psfrag{d3}[c][c][0.7]{$\mathbf{d_3}$}
  \psfrag{d4}[c][r][0.7]{\hspace{0.5cm}$\mathbf{d_4} > 0$}
  \includegraphics*[width=0.3\columnwidth]{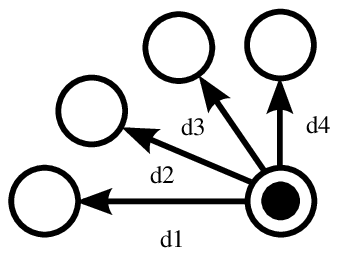} \hspace{1cm}
  \psfrag{d4}[c][r][0.7]{\hspace{0.5cm}$\mathbf{d_4} = 0$}
  \includegraphics*[width=0.3\columnwidth]{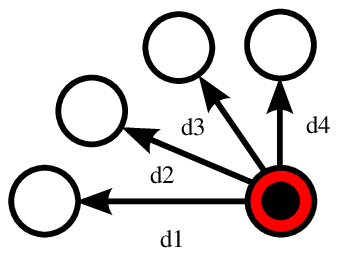}
\caption{\label{openvsclosed} (Color online) Schematic representation of two
  cascades of recurrences of size 4. In both cases $d_4 < d_3 < d_2 <
  d_1$. The cascade on the left is ``open'', i.e. it is possible for a
  fifth recurrence to occur, since $d_4>0$. This is not the case for 
  the cascade on the right. Since $d_4=0$, this cascade is ``closed'' 
  and no distance will be able to break the record.}
\end{figure}

These observations illustrate one of the main features of the OFC model,
namely the repetitive occurrence of events originating at the same
location. Obviously, there is a finite probability that the epicenters
of two events are identical since the OFC model is defined on a discrete
and finite lattice. If this happens, the cascade of recurrences of the
first event ends or is ``closed'' and the number of recurrences of this
event does not increase further (see Fig.~\ref{openvsclosed}) . If this
were exclusively due to the finiteness of the discrete lattice, one
would expect this to happen according to a geometric distribution with
average $L_\text{bulk}^2$. In particular, this effect would not be
relevant in the thermodynamic limit $L_\text{bulk} \to \infty$. However,
our simulations show that the repeated occurrence of the same epicenter
happens much more frequently in the OFC model. In particular, the OFC
catalogs for $s_\text{th}=300$ in Table~\ref{num_events} have $79\%$,
$87\%$ and $92\%$ closed cascades for $\alpha=0.22, 0.20, 0.18$,
respectively
\footnote{The ratio of open vs. closed cascades of
  recurrences, after some initial fluctuations, remains stable as the
  catalog grows, at least for the lengths of time studied. However, since
  the size of the system is finite, eventually the number of closed
  cascades should overcome the number of open ones. The observed stability
  shows that the obtained catalogs are still not affected by this finite
  size effect.}.
This is due to marginal synchronization of neighboring
sites \cite{middleton95} and the existence of quasi-periodic patterns in
the dynamics of the OFC model~\cite{peixoto06} causing the repeated
triggering of epicenters and similar events. Consequently, the average
out-degree of networks generated by the OFC model grows rather
slowly. Moreover, the growth in $\langle k \rangle$ basically saturates
after $\sim 10^3$ events as Fig.~\ref{growth} shows.  This is at least
two orders of magnitude smaller than the rough estimate of
$L_\text{bulk}^2=6.4\times10^5$ based on the lattice size.

When the catalog is shuffled it does not entirely remove the
effect. This is because the average frequency of a given epicenter
does not change, and the distribution of these frequencies is quite
broad, resembling a power law~\cite{peixoto06}. Yet, the shuffling
destroys the relatively short periods of quasi-periodic dynamics
and increases in most cases the time it takes for the activity to
return to the same epicenter and to close a given cascade of
recurrences. Thus, $\left<k\right>$ is larger and its growth is
very close to the predicted behavior for random CST
events until the effects of the broad distribution of the
epicenters frequencies become important after $\sim 10^4$
events.



The network growth described above differs significantly from what is
obtained for seismicity, since real epicenters happen on a continuous
space and therefore never occur exactly in the same place (save, of
course, for limitations in precisely locating the epicenters). 
Hence, the network growth for seismicity --- at least for the given
catalog sizes --- does not
exhibit the ``saturation'' of the average degree obtained for the OFC
model, and grows instead continuously as $\left<k\right>\approx 0.84
\ln(N)$~\cite{davidsen06pm}.

\subsubsection{Degree distributions}

\begin{figure}[htbp]
\psfrag{P(k)}[bc][bc]{$P(k)$}
\psfrag{k}[bc][bc]{$k$}
\psfrag{in, a=0.18}[l][l][0.7]{in, $\alpha=0.18$}
\psfrag{in, a=0.20}[l][l][0.7]{in, $\alpha=0.20$}
\psfrag{in, a=0.22}[l][l][0.7]{in, $\alpha=0.22$}
\psfrag{out, a=0.18}[l][l][0.7]{out, $\alpha=0.18$}
\psfrag{out, a=0.20}[l][l][0.7]{out, $\alpha=0.20$}
\psfrag{out, a=0.22}[l][l][0.7]{out, $\alpha=0.22$}
\includegraphics*[width=\columnwidth]{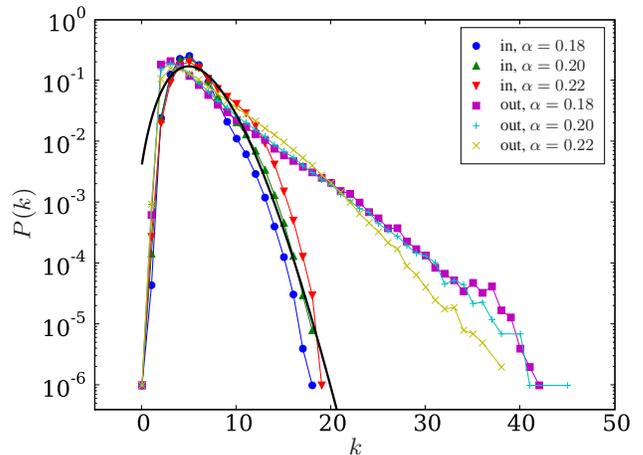}
\caption{\label{topology} (Color online) In-degree and out-degree distributions
  for networks generated by the OFC model for $s_{\text{th}}=300$
  and different $\alpha$'s (see Table~\ref{num_events} for details).
  The solid (black) line is a Poisson distribution with the same
  $\langle k \rangle$ as for $\alpha=0.20$.}
\end{figure}

\begin{figure}[htbp]
\psfrag{P(k)}[bc][bc]{$P(k)$}
\psfrag{k}[bc][bc]{$k$}
\psfrag{in, a=0.18}[l][l][0.7]{in, $\alpha=0.18$}
\psfrag{in, a=0.20}[l][l][0.7]{in, $\alpha=0.20$}
\psfrag{in, a=0.22}[l][l][0.7]{in, $\alpha=0.22$}
\psfrag{out, a=0.18}[l][l][0.7]{out, $\alpha=0.18$}
\psfrag{out, a=0.20}[l][l][0.7]{out, $\alpha=0.20$}
\psfrag{out, a=0.22}[l][l][0.7]{out, $\alpha=0.22$}
\includegraphics*[width=\columnwidth]{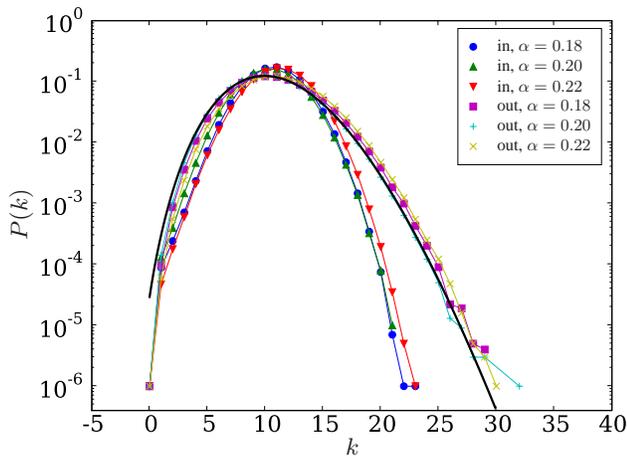}
\caption{\label{topology-shuffled} (Color online) In-degree and
out-degree distributions for networks generated from shuffled
OFC catalogs with $s_{\text{th}}=300$
and different $\alpha$'s. The solid (black) line is a Poisson
distribution with the same $\langle k \rangle$ as for $\alpha=0.20$.}
\end{figure}

A more detailed aspect of the network topology is the distribution of
in- and out-degrees, $P(k_\text{in})$ and $P(k_\text{out})$. In the context of recurrence networks, the out-degree
of a vertex corresponds to its number of recurrences while the in-degree
is the number of events of which it is a recurrence. 
As shown in Fig.~\ref{topology}, the out-degree distribution
deviates significantly from the random CST case, where a Poisson
distribution is expected both for the in- and out-degree
distributions \cite{davidsen06pm}. The out-degree distribution seems instead
to decay exponentially, with inclination slightly dependent on
$\alpha$. The in-degree distribution is closer to a Poissonian. The
overall shape of the distributions also does not depend on the choice of
the lower size threshold for $300 \leq s_{\text{th}} \leq 30000$ (not
shown).
%
%
When the catalogs are shuffled, the obtained out-degree
distribution is very close to a Poissonian (Fig.~\ref{topology-shuffled})
as expected for the random CST case.

The in- and out-degree distributions obtained for the original OFC catalogs are
basically identical to those for seismicity~\cite{davidsen06pm}. 
The only significant differences are the values of $P(1)$ for the 
out-degree. For seismicity, it is much more likely to have an event 
with out-degree one or equivalently an event with a single recurrence.
This is related to the finite extension of an earthquake, as characterized
by its ``rupture length'', and the corresponding microscopic dynamics. 
Thus, our findings suggest the absence of such a length in the OFC model.
This is indeed confirmed by other, more direct means discussed
in Sec.~\ref{spatial}.
Nevertheless, the close resemblance of the degree distributions
for the OFC catalogs and earthquake catalogs suggests that the OFC model
is able to reproduce all other non-trivial gross features of the 
spatiotemporal clustering of seismicity which are captured by the 
degree distributions.

\subsubsection{Degree-degree correlations}

\begin{figure}[htbp]
  \psfrag{<k_{in}>}[bc][bc]{$\left<k^\text{nn}_\text{in}\right>$}
  \psfrag{k_{in}}[bc][bc]{$k_\text{in}$}
  \psfrag{<k_{out}>}[bc][bc]{$\left<k^\text{nn}_\text{out}\right>$}
  \psfrag{k_{out}}[bc][bc]{$k_\text{out}$}
  \includegraphics*[width=\columnwidth]{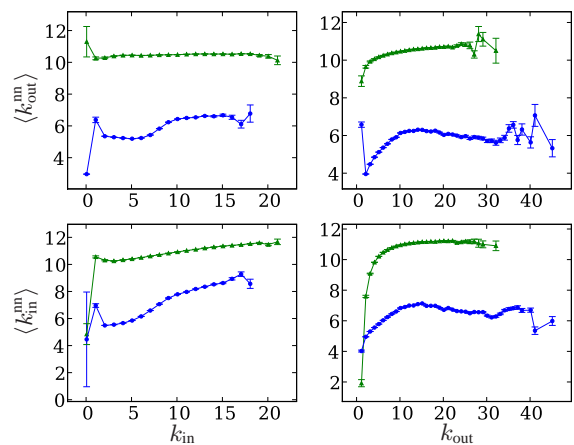}
  \caption{\label{deg-corr} (Color online) Degree-degree correlations
    for $\alpha=0.2$ and $s_{\text{th}}=300$. The curves on the bottom
    (green) are for the original catalog, and on top (blue) for the
    shuffled catalog.  Note that the offset between the curves for the
    original OFC catalog and the shuffled version is simply due to
    different mean degrees.}
\end{figure}

\begin{figure}[htbp]
  \psfrag{<k_{in}>}[bc][bc]{$\left<k_\text{in}\right>$}
  \psfrag{k_{in}}[bc][bc]{$k_\text{in}$}
  \psfrag{<k_{out}>}[bc][bc]{$\left<k_\text{out}\right>$}
  \psfrag{k_{out}}[bc][bc]{$k_\text{out}$}
  \includegraphics*[width=\columnwidth]{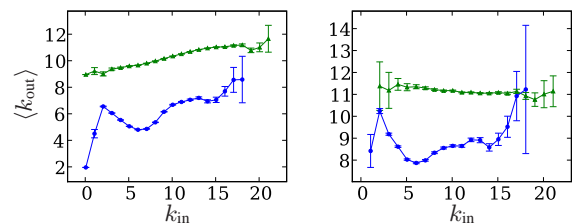}
  \caption{\label{combined-deg-corr}(Color online) Single vertex
    degree-degree correlations for $\alpha=0.2$ and
    $s_{\text{th}}=300$. The left panel shows the results for the
    original catalogs, and the right panel shows the results for the open
    cascade of recurrences only. The bottom curves (green) are for the
    unmodified catalogs, and the top ones (blue) are for the shuffled catalog.}
\end{figure}

The network topology can be further characterized by the
degree-degree correlations~\cite{newman03}. These correlations indicate 
how likely vertices of a given in- or out-degree are connected to 
vertices with another given $k_\text{in}$ or $k_\text{out}$.
We also consider the correlations between $k_\text{in}$ and 
$k_\text{out}$ of the same vertex, the single vertex
degree-degree correlations. To capture both types of correlations, 
we measure the average $k_\text{in}$ or $k_\text{out}$ (either of 
the neighbors or of the same vertex) where the average is taken 
over all vertices with a given $k_\text{in}$ and $k_\text{out}$.
For the OFC catalogs, Figs.~\ref{deg-corr} and~\ref{combined-deg-corr}
show the absence of pronounced non-trivial (single vertex) 
degree-degree correlations.
There are only small qualitative differences between the shuffled and 
unshuffled catalogs
\footnote{The distributions are even independent of 
$\alpha$ or $s_\text{th}$.}
indicating that the degree-degree correlations are not particularly sensitive 
to the given form of spatiotemporal clustering. This is in agreement 
with findings for seismicity \cite{davidsen06pm}.
Interestingly, the single vertex degree-degree correlation function 
$\left<k_\text{out}\right>(k_\text{in})$ increases with $k_\text{in}$
(Fig.~\ref{combined-deg-corr})
--- both for the shuffled and the original OFC catalogs.
This is different from shuffled earthquake catalogs and especially 
from random CST processes which show a slight decrease related to the age
of a vertex~\cite{davidsen06pm}. 
The deviating behavior of the (shuffled) OFC catalogs is due to the closing of 
cascades of recurrences
mentioned above which largely reduces the correlation between the age 
of a vertex 
and its in- and out-degree. Instead, vertices with large $k_\text{in}$ 
correspond to sites with low activity which do not belong to the quasi-periodic
attractors. These sites do not experience the aforementioned
closing of the cascade of recurrences, and thus have a larger
out-degree. The shuffling of the catalog does not alter 
this since it does not affect the closing 
of cascades --- only the time it takes to close. This reasoning is
confirmed by the right panel of Fig.~\ref{combined-deg-corr}, which shows
the single vertex degree correlation, when only open cascades of
recurrences are considered, and the increase is less evident --- and
disappears completely for the shuffled case.

\subsubsection{Clustering coefficient}

The clustering coefficient~\cite{newman03} is another quantity that 
characterizes the network topology. It quantifies how likely neighbors of the
same vertex are connected with each other. In the present context, 
this refers to the probability that
recurrences of the same event are also recurrences of each other. 

For all vertices $i$ with out-degree larger than one, the local clustering
coefficient $C_i$ is given by the ratio of existing links $E_i$
between its $k_i^\text{out}$ recurrences to a maximum possible number of such
links, $\frac{1}{2} k_i^\text{out} (k_i^\text{out} - 1)$. 
The clustering coefficient $C$ of the
network is defined as the average over all vertices $i$ with out-degree
larger than one

\begin{equation}
\label{eq:clustering}
C = \left<C_i\right>_i =  \left< \frac{2E_i}{k_i^\text{out}
  (k_i^\text{out} - 1)} \right>_i.
\end{equation}

The values of the clustering coefficient for the different
OFC catalogs are given in
table~\ref{clustering}. The original catalogs generate
networks which are significantly more clustered than those generated
by the shuffled catalogs. This effect decreases slightly for
larger values of $\alpha$. A substantial amount of clustering seems to
be related to the quasi-periodic attractors of the dynamics generating
vertices with closed cascades of recurrences. This follows directly from
the observation that the
clustering is much less pronounced for those vertices with open cascades
of recurrences (Table~\ref{clustering}). However, the latter
events still retain a higher clustering than the networks for the shuffled
OFC catalogs.

While a direct quantitative comparison with seismicity is not possible
due to different catalog sizes and different mean degrees, 
the qualitative features are the same.
In both cases, the catalogs show a much higher clustering compared to
their respective shuffled counterpart.

\begin{table}
    \caption{\label{clustering} Clustering coefficients
      (Eq.(\ref{eq:clustering})) of the recurrence networks generated by the OFC
      model for different values of $\alpha$ and lower size threshold
      $s_{\text{th}}=300$. The values of $C$ excluding events with
closed cascade of recurrences and for the shuffled OFC catalogs are shown 
for comparison.}
    \begin{ruledtabular}
    \begin{tabular}{c|r|r|r}
      $\alpha$  & $C$         & $C$ (open)  & $C$ (shuffled) \\ \hline
      $0.18$    & $0.4153(8)$ & $0.2064(8)$ & $0.15411(8)$ \\ 
      $0.20$    & $0.3910(3)$ & $0.1970(6)$ & $0.15891(8)$ \\
      $0.22$    & $0.3470(3)$ & $0.1891(4)$ & $0.14650(7)$
    \end{tabular}
    \end{ruledtabular}
\end{table}

\subsection{Temporal distances of recurrences\label{temporal}}

\begin{figure}[htbp]
\psfrag{t}[bc][bc]{$t$}
\psfrag{P(t)}[bc][bc]{$p^{s_{\text{th}}}(t)$} 
\psfrag{P(t)/(N*t^{-1})}[bc][bc][0.7]{$p^{s_{\text{th}}}(t)/(Nt^{-1})$}
\psfrag{t x N}[bc][bc][0.7]{$t \times N$}
\psfrag{s=300}[bc][bc][0.7]{$s_{\text{th}}=300$}
\psfrag{s=500}[bc][bc][0.7]{$s_{\text{th}}=500$}
\psfrag{s=1000}[bc][bc][0.7]{$s_{\text{th}}=1000$}
\psfrag{s=3000}[bc][bc][0.7]{$s_{\text{th}}=3000$}
\psfrag{s=5000}[bc][bc][0.7]{$s_{\text{th}}=5000$}
\psfrag{s=10000}[bc][bc][0.7]{$s_{\text{th}}=10000$}
\psfrag{s=30000}[bc][bc][0.7]{$s_{\text{th}}=30000$}
\includegraphics*[width=\columnwidth]{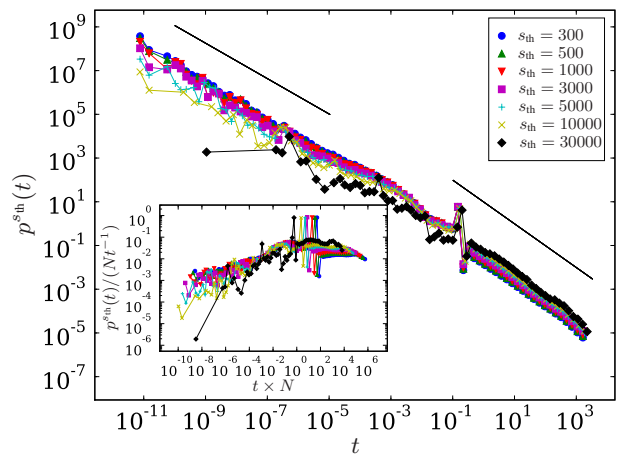}
\caption{\label{delta-iter}(Color online) Distribution of the waiting times
  between events and their recurrences for different threshold sizes
  $s_{\text{th}}$, for $\alpha=0.2$. The solid line on the left is a power law
  with exponent $-0.9$ and on the right one with exponent $-1$.  Inset: Rescaled
  distributions with $N$ taken from table~\ref{num_events}.}
\end{figure}

\begin{figure}[htbp]
\psfrag{P(t)}[bc][bc]{$p^{300}(t)$}
\psfrag{P(t)s}[bc][bc][0.6]{$p^{300}(t)\times (1-4\alpha)$}
\psfrag{t}[bc][bc]{$t$}
\psfrag{t/(1-4\\alpha)}[bc][bc][0.6]{$t/(1-4\alpha)$}
\psfrag{a=0.18}[l][l][0.7]{\hspace{-0.2cm}$\alpha=0.18$}
\psfrag{a=0.20}[l][l][0.7]{\hspace{-0.2cm}$\alpha=0.2$}
\psfrag{a=0.22}[l][l][0.7]{\hspace{-0.2cm}$\alpha=0.22$}
\includegraphics*[width=\columnwidth]{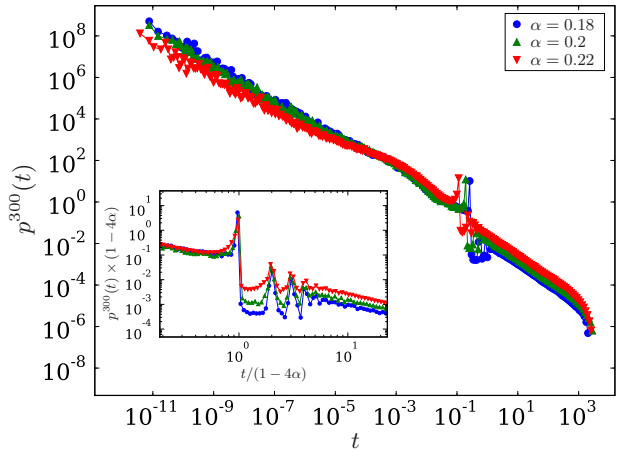}
\caption{\label{delta-iter-alpha}(Color online) Distribution of the
  waiting times between events and their recurrences for
  $s_{\text{th}}=300$ and varying $\alpha$. Inset: Rescaled
  distribution, zoomed around the characteristic time region.}
\end{figure}

The probability density function (PDF) $p^{s_{\text{th}}}(t)$ for the
time intervals or waiting times $t$ associated with the recurrences or
the edges of the network is another important characteristic.
Fig.~\ref{delta-iter} shows this PDF for different OFC catalogs with
$\alpha=0.2$. Despite some variation with $s_{\text{th}}$, 
all PDFs decay approximately as $1/t$ up to the largest 
possible waiting time, determined by the finite time span of the catalog,
with a superposed peak at a characteristic time $t_c$. As the inset 
of Fig.~\ref{delta-iter} shows, this characteristic time separates
two slightly different regimes. While the PDFs decay almost exactly 
as $1/t$ for $t>t_c$, the decay is rather $1/t^{0.9}$ for $t<t_c$.
Note also that $t_c$ does
not depend on $s_{\text{th}}$. Such a characteristic time is expected
since the activity shows quasi-periodic behavior as discussed in
Section~\ref{sec:growth}. Moreover, $t_c$ depends on
the dissipation parameter $\alpha$, as Fig.~\ref{delta-iter-alpha}
shows. More specifically (as the inset of Fig.~\ref{delta-iter-alpha}
shows), the characteristic time seems to be exactly 
$1-4\alpha$, which is the natural period of the system:
For periodic boundary conditions and relatively small $\alpha$, 
the system exhibits only trivial, 
perfectly synchronized topplings of size one and the activity returns 
to each site with period $1-4\alpha$~\cite{grassberger_efficient_1994}. 
It is even possible to identify secondary characteristic recurrence 
times in the inset of Fig.~\ref{delta-iter-alpha}, corresponding to 
integer multiples of $1-4\alpha$, again emphasizing the 
quasi-periodic behavior
\footnote{ 
The shape of $p^{s_{\text{th}}}(t)$ does not change if recurrences
belonging to closed cascades are excluded. In particular,
the characteristic time scale $t_c$ is still present though slightly
less pronounced.}.

\begin{figure}[htbp]
\psfrag{t}[bc][bc]{$t$}
\psfrag{P(t)}[bc][bc]{$p^{s_{\text{th}}}(t)$} 
\psfrag{P(t)/(N*t^{-1})}[bc][bc][0.7]{$p^{s_{\text{th}}}(t)/(Nt^{-1})$}
\psfrag{t x N}[bc][bc][0.7]{$t \times N$}
\psfrag{s=300}[bc][bc][0.7]{$s_{\text{th}}=300$}
\psfrag{s=500}[bc][bc][0.7]{$s_{\text{th}}=500$}
\psfrag{s=1000}[bc][bc][0.7]{$s_{\text{th}}=1000$}
\psfrag{s=3000}[bc][bc][0.7]{$s_{\text{th}}=3000$}
\psfrag{s=5000}[bc][bc][0.7]{$s_{\text{th}}=5000$}
\psfrag{s=10000}[bc][bc][0.7]{$s_{\text{th}}=10000$}
\psfrag{s=30000}[bc][bc][0.7]{$s_{\text{th}}=30000$}
\includegraphics*[width=\columnwidth]{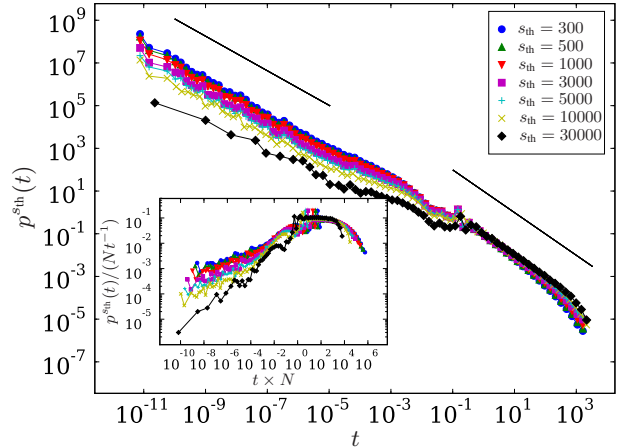}
\caption{\label{delta-iter-shuffled}(Color online) Distribution of waiting times
  between events and their recurrences for shuffled OFC catalogs with different
  threshold sizes $s_{\text{th}}$ and $\alpha=0.2$. The solid line on the left
  is a power law with exponent $-0.9$ and on the right one with exponent $-1$.
  Inset: Rescaled distribution with $N$ taken from table~\ref{num_events}.}
\end{figure}

For the shuffled OFC catalogs, $p^{s_{\text{th}}}(t)$ is shown in
Fig.~\ref{delta-iter-shuffled}.  As expected, the peak at the
characteristic time scale $t_c$ is barely noticeable since the quasi-periodic
behavior is basically absent in the shuffled catalog. Yet, all other
qualitative features are preserved.
In particular, the PDF is very different from the behavior of 
$p^{s_{\text{th}}}(t)$ in the random CST case where a flat regime for small
$t$ and $1/t$ decay for large $t$ exist~\cite{davidsen06pm}.
In the latter case, the transition time $t^*$ between the two 
regimes scales with the inverse of the rate of events $t^* \sim T/N$. 
We do not observe such a scaling for the shuffled OFC catalogs as follows
from the inset of Fig.~\ref{delta-iter-shuffled}. While the data do collapse
for large arguments, there is no good collapse for small arguments. 
This is due to the fact that there is a non-negligible probability that
the time interval between subsequent events in the OFC model is 
arbitrarily small~\cite{davidsen00,drossel_complex_2002} --- 
in sharp contrast to the assumption of a Poisson process in the random CST 
case.

For seismicity $t^*$ --- and indeed $p^{s_{\text{th}}}(t)$ --- 
does not depend on $N$ (corresponding to a varying magnitude threshold) 
and there is also no equivalent to $t_c$~\cite{davidsen06pm}. These
observations clearly prove that the OFC model has certain features 
which are incompatible with seismicity and which strongly influence 
the generated spatiotemporal clustering of events.

\subsubsection{Temporal hierarchy of subsequent recurrences}

\begin{figure}[htbp]
\psfrag{t_i/t_{i+1}}[bc][bc][0.7]{$t_i/t_{i+1}$}
\psfrag{p(t_i/t_{i+1})}[bc][bc][0.7]{$p(t_i/t_{i+1})$}
\psfrag{i=0}[c][c][0.5]{$\mathbf{i=1}$} 
\psfrag{i=1}[c][c][0.5]{$\mathbf{i=2}$}
\psfrag{i=2}[c][c][0.5]{$\mathbf{i=3}$}
\psfrag{i=4}[c][c][0.5]{$\mathbf{i=5}$}
\psfrag{i=6}[c][c][0.5]{$\mathbf{i=7}$}
\includegraphics*[width=1.1\columnwidth]{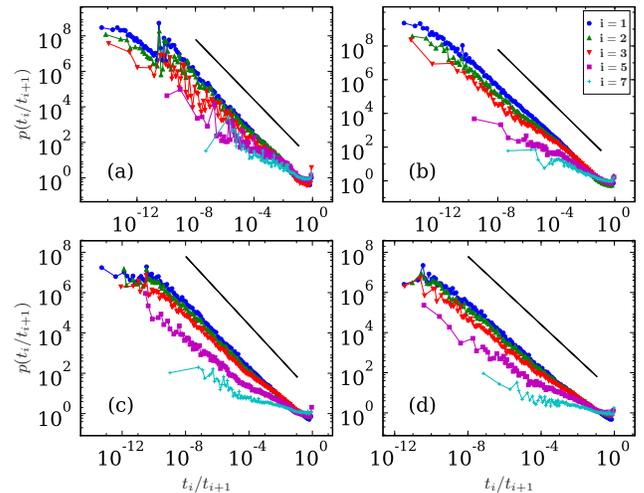}
\caption{\label{delta-iter-ratio}Distribution of the waiting time ratios 
  for subsequent recurrences, for $\alpha=0.2$ and
  $s_\text{th}=300$, for (a) the regular catalog, (b) only the
  ``open'' recurrences, (c) the shuffled catalog and (d) only the
  ``open'' recurrences for the shuffled catalog. In all plots the
  solid line is a power-law with inclination $-0.85$.}
\end{figure}

Another important aspect of the network of recurrences is the possible
existence of hierarchies of recurrences. This can be captured by the
ratios of the waiting times $t_i/t_{i+1}$ for 
subsequent recurrences belonging to the same originating
event. Here, it is assumed that the recurrences for a given event are
ordered according to their time of occurrence such that the $i$-th
recurrence is followed by the $i+1$-th recurrence for all ranks
$i$. The corresponding PDFs are shown in
Fig.~\ref{delta-iter-ratio}. For the OFC catalogs, panel (a) indicates
that the PDFs vary with $i$. While they closely resemble a power law 
with exponent $-0.85$ in an intermediate regime for small $i$, 
deviations occur for 
larger $i$'s. This is independent of $\alpha$ and $s_{\text{th}}$ (not shown).
Independent of $i$, there is also a peak around $t_i/t_{i+1}=1$. As
Fig.~\ref{delta-iter-ratio}~(c) shows, there are only few discernible
differences between the original and the shuffled catalogs: The
peak is slightly more pronounced in the 
original catalog implying that subsequent recurrences are more likely 
to be separated by a very short time interval only, and the occurrence 
of smaller ratios for larger values of $i$ is more likely for the 
shuffled catalogs. A third and more pronounced but expected difference 
is that the 
kink related to $t_C$ around $t_i/t_{i+1} \sim 10^{-10}$ in the original 
catalog is absent in the shuffled catalog. 

Even these small differences in the PDFs disappear if only open 
cascade of recurrences are considered, as shown in 
Fig.~\ref{delta-iter-ratio}~(b) and (d). In all cases the PDFs depend 
very weakly on the value of $\alpha$, with all
mentioned properties equally present in the range of $\alpha$'s studied.


For seismicity, the ratios $t_i/t_{i+1}$ for subsequent recurrences
revealed a different non-trivial characteristic~\cite{davidsen06pm}. 
Their distribution showed two power-law regimes --- in contrast to a 
single power law for the shuffled catalog. The
spatiotemporal correlations responsible for these results do not seem
to be present in the OFC model, as the above analysis shows.

\subsection{Spatial distances of recurrences\label{spatial}}

\begin{figure}[htbp]
\psfrag{P(l)}[bc][bc]{$p^{s_\text{th}}(l)$}
\psfrag{l}[bc][bc]{$l$}
\psfrag{F(l/s^\{0.30\}; p=1.50)}[bc][bc][0.7]{$F(l/s_\text{th}^{0.3})$}
\psfrag{l/s^\{0.3\}}[bc][bc][0.7]{$l/s_\text{th}^{0.3}$}
\psfrag{s=300}[bc][bc][0.7]{$s_{\text{th}}=300$}
\psfrag{s=500}[bc][bc][0.7]{$s_{\text{th}}=500$}
\psfrag{s=1000}[bc][bc][0.7]{$s_{\text{th}}=1000$}
\psfrag{s=3000}[bc][bc][0.7]{$s_{\text{th}}=3000$}
\psfrag{s=5000}[bc][bc][0.7]{$s_{\text{th}}=5000$}
\psfrag{s=10000}[bc][bc][0.7]{$s_{\text{th}}=10000$}
\psfrag{s=30000}[bc][bc][0.7]{$s_{\text{th}}=30000$}
\includegraphics*[width=\columnwidth]{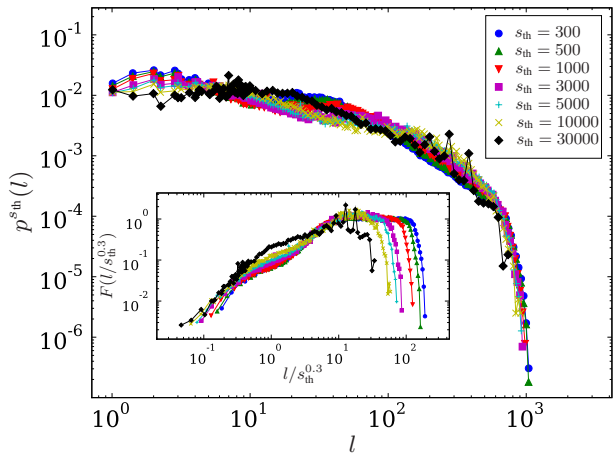}
\caption{\label{recurrence-dist} (Color online) Distribution of
  recurrence distances for $\alpha=0.2$ and different threshold values $s_\text{th}$.
  Inset: Distributions rescaled according to Eq.(\ref{scaling_ansatz})
  with $\delta=0.3$ and $\beta=1.5$.
  Recurrences of distance $l=0$ are
  not shown, and were not used in the normalization.}
\end{figure}

\begin{figure}[htbp]
\psfrag{P(l)}[bc][bc]{$p^{s_\text{th}}(l)$}
\psfrag{l}[bc][bc]{$l$}
\psfrag{F(l/s^\{0.30\}; p=1.35)}[bc][bc][0.7]{$F(l/s_\text{th}^{0.3})$}
\psfrag{l/s^\{0.3\}}[bc][bc][0.7]{$l/s_\text{th}^{0.3}$}
\psfrag{s=300}[bc][bc][0.7]{$s_{\text{th}}=300$}
\psfrag{s=500}[bc][bc][0.7]{$s_{\text{th}}=500$}
\psfrag{s=1000}[bc][bc][0.7]{$s_{\text{th}}=1000$}
\psfrag{s=3000}[bc][bc][0.7]{$s_{\text{th}}=3000$}
\psfrag{s=5000}[bc][bc][0.7]{$s_{\text{th}}=5000$}
\psfrag{s=10000}[bc][bc][0.7]{$s_{\text{th}}=10000$}
\psfrag{s=30000}[bc][bc][0.7]{$s_{\text{th}}=30000$}
\includegraphics*[width=\columnwidth]{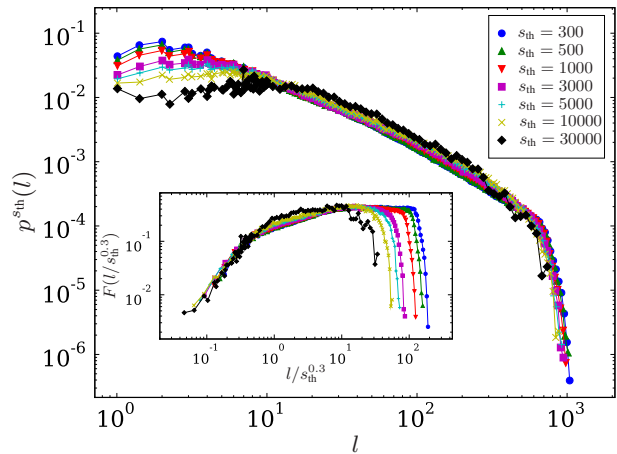}
\caption{\label{recurrence-dist-open} (Color online) Distribution of
  recurrence distances for $\alpha=0.2$ and different threshold values $s_\text{th}$,
  but only considering ``open'' cascades of recurrences. Inset: Distributions
  rescaled according to Eq.(\ref{scaling_ansatz}) with $\delta=0.3$ and $\beta=1.35$.}
\end{figure}

Exactly like the waiting times of recurrences, the spatial
distances between an event and its recurrences can be obtained for the
OFC model for further characterization and comparison with seismicity.
The PDF of these distances
$p^{s_{\text{th}}}(l)$ is shown in Fig.~\ref{recurrence-dist} for
$\alpha=0.2$ and different size thresholds. The results for other 
values of $\alpha$ are qualitatively similar. The overall shape
of the distributions is rather broad with cut-offs at short and 
long distances due to the discrete and finite lattice, respectively,
and varies with $s_{\text{th}}$. Yet, for arguments less than the 
long distance cut-off two different regimes seem to be generally 
present. The inset of Fig.~\ref{recurrence-dist} shows an
attempted scaling collapse according to the ansatz
\begin{equation}
\label{scaling_ansatz}
 p^{s_{\text{th}}}(l) = \frac{(l/s_{\text{th}}^{\delta})^{-\beta}F(l/s_{\text{th}}^{\delta})}{s_{\text{th}}^{\delta}},
\end{equation}
with $\delta=0.3$ and $\beta=1.5$. If successful for arguments 
significantly less than the respective finite size cut-off, it would 
imply the existence of a characteristic
distance dependent on $s_{\text{th}}$ with $l^*\sim
s_{\text{th}}^{0.3}$ --- qualitatively similar to seismicity and 
the random CST case as discussed below. However, as the inset of
Fig.~\ref{recurrence-dist} clearly shows, the data do not collapse
as intended and a collapse does not seem to be possible at all. When
only open cascades of recurrences are considered, as shown in
Fig.~\ref{recurrence-dist-open}, the PDFs are slightly different but 
there is no convincing collapse either. It does, however, show again that the
statistics of those events that have closed cascades and predominantly
belong to quasi-periodic attractors, are significantly different 
from the ``open'' ones.

\begin{figure}[htbp]
\psfrag{P(l)}[bc][bc]{$p^{s_\text{th}}(l)$}
\psfrag{l}[bc][bc]{$l$}
\psfrag{s=300}[bc][bc][0.7]{$s_{\text{th}}=300$}
\psfrag{s=500}[bc][bc][0.7]{$s_{\text{th}}=500$}
\psfrag{s=1000}[bc][bc][0.7]{$s_{\text{th}}=1000$}
\psfrag{s=3000}[bc][bc][0.7]{$s_{\text{th}}=3000$}
\psfrag{s=5000}[bc][bc][0.7]{$s_{\text{th}}=5000$}
\psfrag{s=10000}[bc][bc][0.7]{$s_{\text{th}}=10000$}
\psfrag{s=30000}[bc][bc][0.7]{$s_{\text{th}}=30000$}
\psfrag{l x N^\{0.18\}}[bc][bc][0.7]{$l \times N^{0.18}$}
\psfrag{P(l x N^\{0.18\})*l^\{1.00\}/N^\{0.18\}}[bc][bc][0.7]{$P(l \times N^{0.18})/(N^{0.18}l^{-1})$}
\includegraphics*[width=\columnwidth]{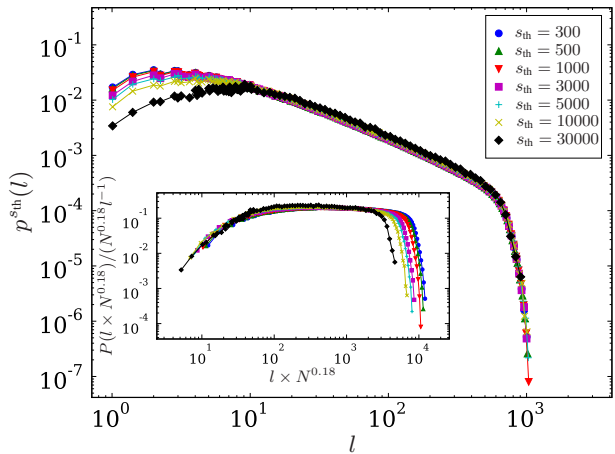}
\caption{\label{recurrence-dist-shuffled}(Color online) Distribution of recurrence distances
  for the shuffled OFC catalog with $\alpha=0.2$ and different threshold values $s_\text{th}$. Inset: Distributions
  rescaled according to the number of events (see Table~\ref{num_events}) as indicated in the legend.}
\end{figure}

For the shuffled OFC catalog the corresponding results for
$p^{s_{\text{th}}}(l)$ are shown in
Fig.~\ref{recurrence-dist-shuffled}. In this case, a collapse is indeed
possible. Here, the collapse was done using the number
of events $N$ of the respective catalog given in
Table~\ref{num_events}. Due to the absence of non-trivial correlations
in the shuffled catalogs, this is 
equivalent to the scaling with $s_{\text{th}}$ since 
$N \propto s_{\text{th}}^{-0.8}$ as follows from
the distribution of event sizes $P(s)\sim s^{-1.8}$.
These findings can be directly compared to the
random CST case for which $l^* \propto N^{-\gamma}$, where $\gamma=1/D$ 
and $D$ is the 
underlying spatial dimension~\cite{davidsen06pm}.  However, for the 
shuffled OFC catalog we find $l^* \propto N^{-0.18}$ which would 
imply $D \approx 5$. 
This is in sharp contrast to the fact that the OFC model is defined
on a 2D lattice. This disagreement can be attributed to the assumption of
continuous space made to derive the results in Ref.~\cite{davidsen06pm}. 
On a discrete lattice, when
the characteristic distance approaches the fundamental lattice
distance, the identification of the exponent with $1/D$ cannot be
expected. Despite this mismatch, the random CST process correctly
predicts the power-law decay with exponent $1$ for large distances 
of the shuffled OFC catalogs. Note that this decay clearly distinguishes
the shuffled OFC catalogs from the original OFC catalogs.

\begin{figure}[htbp]
  \psfrag{P(l)/N^g}[bc][bc]{$p^{300}(l)/\Delta t^\gamma$}
  \psfrag{l x N^g}[bc][bc]{$l \times \Delta t^\gamma$}

  \psfrag{t=154}[l][l][0.6]{\hspace{-0.4cm}$\Delta t = 154$}
  \psfrag{t=309}[l][l][0.6]{\hspace{-0.4cm}$\Delta t = 309$}
  \psfrag{t=463}[l][l][0.6]{\hspace{-0.4cm}$\Delta t = 463$}
  \psfrag{t=617}[l][l][0.6]{\hspace{-0.4cm}$\Delta t = 617$}
  \psfrag{t=1235}[l][l][0.6]{\hspace{-0.4cm}$\Delta t = 1235$}
  \psfrag{t=1852}[l][l][0.6]{\hspace{-0.4cm}$\Delta t = 1852$}
  \psfrag{t=2469}[l][l][0.6]{\hspace{-0.4cm}$\Delta t = 2469$}
  \psfrag{t=3086}[l][l][0.6]{\hspace{-0.4cm}$\Delta t = 3086$}

  \includegraphics*[width=1.1\columnwidth]{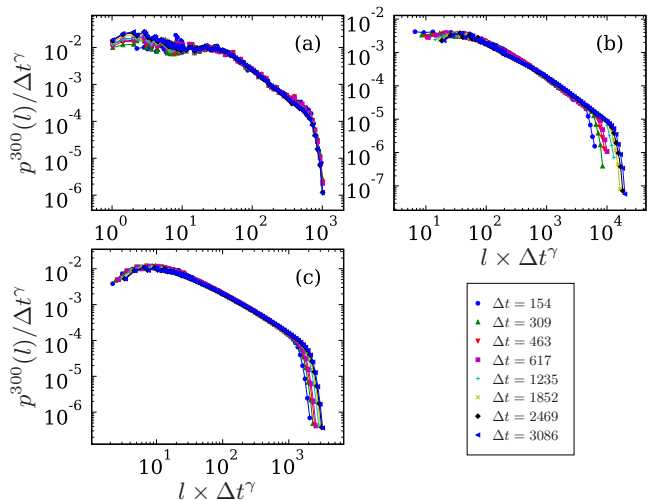}
  \caption{\label{recurrence-dist-parts-scaled}(Color online)
    Distribution of recurrence distances for $\alpha=0.2$ and
    $s_{\text{th}}=300$, for an increasingly larger potion of the
    catalog $\Delta t$, re-scaled as indicated in the legend, for (a)
    the regular catalog ($\gamma=0$, i.e. no scaling), (b) only the
    ``open'' recurrences ($\gamma=0.37$) and (c) the shuffled catalog
    ($\gamma=0.18$).}
\end{figure}

The equivalence of the scaling of $p^{s_\text{th}}(l)$ with
$s_\text{th}$ and with $N$ for the shuffled OFC catalogs discussed above
is confirmed by Fig.~\ref{recurrence-dist-parts-scaled} (c). There, the
variation of $p^{s_\text{th}}(l)$ with the time length $\Delta t$ of the
catalog is explicitly shown for $s_\text{th}=300$.  Note that the x and
y axis are rescaled with $\Delta t^\gamma$, where $\Delta t$ is the time length
indicated in the legend.  Not only the shuffled catalog varies with time
(and hence $N$) for small distances but also the original catalog --- even if
only open cascades of recurrences are considered 
(see Fig.~\ref{recurrence-dist-parts-scaled} (a) and (b)).
In the latter case, the different curves can be collapsed using the same
ansatz as for the shuffled catalog with $\gamma=0.37$, while the time
dependence of the unmodified catalog seems to be more elaborate (it is
shown unscaled in Fig.~\ref{recurrence-dist-parts-scaled}). Moreover,
comparing Figs.~\ref{recurrence-dist-parts-scaled} (b) and the inset of
\ref{recurrence-dist-open} allows to test the scaling relation $N^\gamma
\propto s_{\text{th}}^{-0.8\gamma}$ in the case of open cascades of 
recurrences. Indeed, we have  $0.8\times0.37 \approx 0.3$ assuming
$N\propto\Delta t$. This would imply that the dependence of
$p^{s_\text{th}}(l)$ on $s_\text{th}$ is simply due to the variation in
$N$. Yet, the fact that $p^{s_\text{th}}(l)$ does not obey scaling
suggests that the situation is more complicated and that there is 
a non-trivial dependence on $s_\text{th}$.

For seismicity, $p^{s_{\text{th}}}(l)$ shows a
characteristic distance $l^*$, after which the distribution decays as
a power-law with an exponent close to one~\cite{davidsen06pm}. 
The value of $l^*$ depends only on the lower magnitude
threshold and, in particular, \emph{does not change with time} (or $N$)
for fixed $m$. More specifically the scaling of this characteristic
distance obeys the ansatz $l^*(m) \propto 10^{0.45 m}$, where $m$ is the
lower threshold magnitude. This allows one to identify $l^*$ with the
rupture length of an earthquake --- its spatial extent --- even though
it is explicitly neglected in the point process description. 
In particular, it
directly relates the underlying microscopic dynamics to the 
statistical properties of the network of recurrences --- one of the main
achievements for seismicity. As shown above, 
the OFC model does not successfully reproduce these features. While the 
characteristic distance does depend on the lower threshold size, 
no simple scaling
ansatz can be identified (Eq.(\ref{scaling_ansatz})), even when only
``open'' cascades of recurrences are considered. More importantly, the 
characteristic distance decreases with time in both cases.

\subsubsection{Spatial hierarchy of subsequent recurrences}

\begin{figure}[htbp]
\psfrag{l_{1+i}/l_i}[bc][bc][0.7]{$l_{i+1}/l_i$}
\psfrag{p(l_{1+i}/l_i)}[bc][bc][0.7]{$p(l_{i+1}/l_i)$}
\psfrag{i=0}[c][c][0.5]{$\mathbf{i=0}$} 
\psfrag{i=1}[c][c][0.5]{$\mathbf{i=1}$}
\psfrag{i=2}[c][c][0.5]{$\mathbf{i=2}$}
\psfrag{i=3}[c][c][0.5]{$\mathbf{i=3}$}
\psfrag{i=4}[c][c][0.5]{$\mathbf{i=4}$}
\includegraphics*[width=1.1\columnwidth]{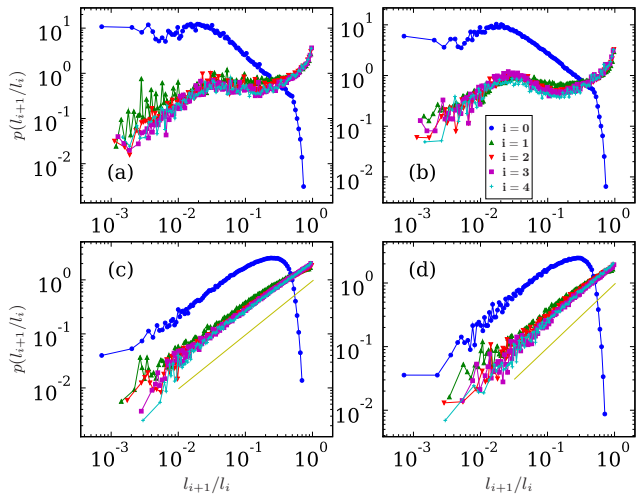}
\caption{\label{recurrence-dist-ratio}Distribution of ratios between
  subsequent recurrence distances for $\alpha=0.2$ and
$s_\text{th}=300$, for
  (a) the regular catalog, (b) only the ``open'' recurrences, (c)
  the shuffled catalog and (d) only the ``open'' recurrences for the
  shuffled catalog. The solid line is a power law with exponent $1$.}
\end{figure}

The spatial hierarchy of recurrences can be captured by the distribution 
of distance ratios $l_{i+1}/l_i$ for subsequent recurrences belonging 
to the same originating event. Here, it is assumed that the recurrences 
for a given event are
ordered according to their time of occurrence such that the $i$-th
recurrence is followed by the $i+1$-th recurrence. 
The corresponding PDFs are shown in
Fig.~\ref{recurrence-dist-ratio} for $\alpha=0.2$ and
$s_\text{th}=300$ but the PDFs seem to be independent of the specific
values of $\alpha$ and $s_{\text{th}}$ in the range we studied. We have 
also included the case $i=0$ for which we use $l_0=\sqrt{2}L_{bulk}$.
Note that $l_1$ corresponds to the distances between subsequent events 
and that in this case
there is a cut-off at large arguments due to the finite lattice size.
For all $i>0$ the PDFs shown in Fig.~\ref{recurrence-dist-ratio} seem to
be independent of $i$. For the original OFC catalog
(Fig.~\ref{recurrence-dist-ratio}(a)), the PDFs increase for
small arguments and then become flat for larger ones, before 
increasing again as the ratio approaches $1$.
The same behavior is observed when only open cascades of recurrences 
are considered (Fig.~\ref{recurrence-dist-ratio}(b)). 

The situation is different for the shuffled OFC catalogs. As shown in
Fig.~\ref{recurrence-dist-ratio}(c), all PDFs increase as a power-law
with exponent $1$. This is similar to the random CST model which
predicts $p(x) = Dx^{D-1}$, independent of $i$, with
$x=l_{i+1}/l_i$~\cite{davidsen06pm}. For the shuffled OFC catalogs, this
would imply a spatial dimension $D=2$ --- exactly as expected. 
The same behavior is observed when only open cascades of recurrences 
are considered (Fig.~\ref{recurrence-dist-ratio}(d)). 
The comparison of the OFC catalogs and their shuffled counterparts
suggests that the distribution of ratios between subsequent recurrence 
distances indeed captures non-trivial aspects of the spatiotemporal dynamics.

All cases are however very different
from seismicity, where the ratios all decay as a power-law with
exponent $\sim 0.6$.

\section{Discussion \& Conclusion\label{summary}}

We have studied the network of recurrences of the OFC model, with the
objective of characterizing its causal structure and comparing it to
seismicity. We have found that, in agreement with previous
results~\cite{hergarten02,helmstetter04,peixoto06}, the model does
exhibit some non-trivial features observed in seismicity, but it comes
short of a complete qualitative description of several robust features
of spatiotemporal clustering captured by the used method. More
specifically, the model reproduces central topological features of
the recurrences network, which are the degree distributions and high 
clustering, as well as the absence of non-trivial degree correlations
and some other partial aspects of the recurrence distance
and time distributions. However, the two most important properties of
the dynamics of seismicity identified by this recurrence approach are,
firstly, that the time (distance) distributions are independent of
magnitude thresholds (time), and secondly, that the distance 
statistics provides an
independent assessment of the rupture length of earthquakes. The OFC
model fails to reproduce the first feature, as both the time and
distance distributions vary with magnitude threshold and time, respectively. 
The second feature is also not
reproduced. While there is a dependence of the distance distribution
on the size of the events, there is no scaling which would allow to 
identify a rupture
length. One third, and more subtle, characteristic of the recurrences
which is not reproduced by the OFC model is the hierarchy of
recurrences, given by the distribution of ratios of time and distance
between consecutive recurrences, which in the case of seismicity
deviates significantly from the random case. Moreover, the whole
analysis exposes another crucial aspect of the OFC model, which is the
elevated occurrence of epicenters at distance zero from one another,
tightly connected to quasi-periodicity and well distinct from
seismicity.

Obviously, the OFC model was not intended to be a complete
description of seismicity. It was mainly proposed as a conceptual
model and a possible origin of the Gutenberg-Richter law from a
SOC point of view. Also, like the Burridge-Knopoff spring-block
model, it aims to model the dynamics of a \emph{single fault}, and not
the complex interplay of different fault structures as observed in
reality. Despite these limitations, the OFC model is capable of
showing a rich dynamics, reproducing even some more subtle
features of the spatiotemporal clustering of earthquakes.

\section{Acknowledgments}

We thank the organizers of the Workshop on ``Dynamics on Complex Networks 
and Application'' at the Max-Planck-Institute for Physics of Complex Systems,
Dresden, Germany, where this work was initiated. We also thank Peter 
Grassberger for his comments on the manuscript.
Part of this work was supported by Funda\c{c}\~{a}o de Amparo \`{a} Pesquisa
do Estado de S\~{a}o Paulo (FAPESP), process number 03/03429-6.

\bibliography{j2}

\end{document}